\documentclass[aps,prl,onecolumn,superscriptaddress,groupedaddress]{revtex4}  
\usepackage{graphicx,comment,float,braket,textcomp,amsmath,lipsum,color,soul}  
\usepackage{dcolumn}   
\usepackage{bm}        
\usepackage{amssymb}   
\newcommand{\e}{e}
\newcommand{\imag}{\text{Im}}

\begin{document}

\title{An Exact Expression for Multidimensional Spectroscopy of a Spin-Boson Hamiltonian}

\author{Albert Liu} \affiliation{Condensed Matter Physics and Materials Science Division, Brookhaven National Laboratory, Upton, New York, 11973 USA}

\vskip 0.25cm

\date{\today}

\begin{abstract}
    Multidimensional coherent spectroscopy is a powerful tool to characterize nonlinear optical response functions. Typically, multidimensional spectra are interpreted via a perturbative framework that straightforwardly provides intuition into the density matrix dynamics that give rise to specific spectral features. When the goal is to characterize system coupling to a thermal bath however, the perturbative formalism becomes unwieldy and yields less intuition. Here, we extend an approach developed by Vagov et al. \cite{Vagov2002} to provide an exact expression for multidimensional spectra of a spin-boson Hamiltonian up to arbitrary order of electric field interaction. We demonstrate the utility of this expression by modeling polaron formation and coherent exciton-phonon coupling in quantum dots, which strongly agree with experiment.
\end{abstract}

\maketitle

\section{Introduction}

Many quantum systems of both fundamental and practical interest may be considered open quantum systems \cite{Breuer_Petruccione2007_Book}, in which relevant degrees of freedom (the `system') are coupled in some significant way to a thermal environment (the `bath'). In recent years, the interaction between the system and the bath has become a topic of increasing interest to varied communities \cite{GardinerandZoller2004_Book,Mohsenietal2014_Book}. This is driven by two primary motivations, namely that system-bath coupling is essential to understanding complex quantum systems \cite{Rotter2015,Mori2023} and that optimizing thermal decoherence is essential for quantum technologies \cite{Reich2015,Man2015,Miller2022}.

Optical spectroscopy is perhaps the most direct way of interrogating open quantum systems. Most commonly, linear spectroscopies such as absorption or fluorescence spectroscopy \cite{Fox2010_Book} return spectral lineshapes that provide insight into coherence dephasing, and thereby system-bath coupling \cite{Fidy1998,Wendling2000}. These methods fail in the presence of disorder however, where an inhomogeneous resonance energy distribution obscures the homogeneous lineshapes. In response to this challenge, nonlinear spectroscopic techniques were developed to extract homogeneous lineshapes even in the presence of dominant disorder. For example, photon echo spectroscopy \cite{Cundiff2008} and spectral hole burning \cite{Moerner1998_Book} are two such nonlinear spectroscopies that have been widely applied to a variety of atomic \cite{Lorenz2005,Hetet2008}, molecular \cite{Cho1992}, and solid-state \cite{Jakubczyk2016,Liu2021_MQT} systems.

Undoubtedly the most general nonlinear spectroscopic technique is multidimensional coherent spectroscopy (MDCS) \cite{Cundiff2013,MDCS_Book}. An optical analogue of nuclear magnetic resonance, MDCS involves impulsive excitation of a system by a sequence of laser pulses and measurement of the resultant nonlinear wave-mixing signal. Dynamics of the system density matrix are encoded in the inter-pulse time-delays \cite{Mukamel1999_Book}, and Fourier transform of the nonlinear signal along two or more of the experimental time axes returns a multidimensional spectrum that contains a wealth of information about the microscopic physics. Multidimensional spectra are typically interpreted perturbatively \cite{Mukamel1999_Book,MDCS_Book}, in which electric field interactions induce sequential changes in the system density matrix, which provides clear intuition for the coherent and incoherent dynamics that evolve across various time axes \cite{Cho2009,HammZanni2012}. However, the formalism becomes cumbersome upon incorporating system-bath coupling, particularly for higher-order nonlinearities \cite{Tanimura1997,Zhang2013}.

Here, we present an exact, non-perturbative expression for multidimensional spectroscopy of a spin-boson Hamiltonian, a ubiquitous model for open quantum systems. Derived from a treatment developed by Vagov et al. \cite{Vagov2002}, this compact expression provides a way to more easily model many complex systems and interpret their spectra. We demonstrate its utility by reproducing experimental signatures of polaron formation and coherent exciton-phonon coupling in quantum dots.

\section{Spin-Boson Hamiltonian}

Specifically, the model we consider is that of a spin-boson Hamiltonian \cite{Leggett1987,Calarco2003,Schlosshauer2007} without tunneling, often termed in the literature as the `independent boson Hamiltonian' \cite{Chenu2019}. This consists of a (electronic) two-level `system', a collection of (vibrational) harmonic oscillators that comprise the `bath', and coupling between the two is retained to lowest order (bi-linearly). For ease of comparison, we use the notation of \cite{Vagov2002} and the Hamiltonian reads:
\begin{align}
    \nonumber H &= \hbar\Omega c^\dagger c + \sum\limits_{\bf q}\hbar\omega_{\bf q} b^\dagger_{\bf q} b_{\bf q} - (MEc^\dagger d^\dagger + M^*E^*dc)\\
    &\hspace{0.9cm} + \sum\limits_{\bf q} \hbar\left[(g^e_{\bf q} b_{\bf q} + g^{e*}_{\bf q} b^\dagger_{\bf q})c^\dagger c - (g^h_{\bf q} b_{\bf q} + g^{h*}_{\bf q} b^\dagger_{\bf q})d^\dagger d\right],
\end{align}
where $c^{(\dagger)}$, $d^{(\dagger)}$, and $b^{(\dagger)}_{\bf q}$ are the (creation) annihilation operators for electrons, holes, and a vibrational mode of wavevector ${\bf q}$ respectively. In order, $\hbar\Omega$ is the electronic energy splitting, $\hbar\omega_{\bf q}$ is the energy of the respective vibrational mode, $M$ is the electronic dipole moment that interacts with an external electric field $E(t)$, and $g^{(e)h}_{\bf q}$ is the coupling strength between the (electrons) holes and each vibrational mode. Note that from a classical perspective the final coupling term shifts the vibrational coordinate of minimum energy upon electronic excitation, which results in the displaced oscillator model for a single mode.


\subsection{Spectral Density}

The effects of system-bath coupling may be captured in a quantity termed the spectral density $C''(\omega_v)$ \cite{Fleming1996,Calarco2003,Nitzan2006}, which resolves the coupling strength as a function of vibrational frequency $\omega_v$. Here, we first consider a Debye contribution \cite{Wang1999} to the spectral density at low frequencies, a common functional form used to model a variety of cases ranging from molecular systems \cite{Kjellberg2006} to quantum dots \cite{Seibt2013}:
\begin{align}\label{Eq_SpectralDensity}
    \nonumber C''_D(\omega_v) &= \sum\limits_{\bf q}|g^e_{\bf q} - g^h_{\bf q}|^2\delta(\omega_v - \omega_{\bf q})\\
    &= 2\pi S_D\frac{\omega_v^4}{2\omega_c^3}\exp\left[-\omega_v/\omega_c\right]
\end{align}
where we use values for the Huang-Rhys factor $S_D = 0.05$ and cutoff frequency $\omega_c = 0.2$ THz.

\subsection{Impulsive Excitation}

Dynamics of the above Hamiltonian in response to external driving are often obtained by perturbative solution of density matrix equations of motion \cite{Butkus2012,Seibt2013}, but the expressions can be unwieldy for nonlinear responses while the dynamics of bath degrees of freedom are obscured. To obtain an exact solution for multidimensional spectroscopy of the spin-boson Hamiltonian described above, we extend a treatment by Vagov et al. \cite{Vagov2002} for impulsive excitation in terms of three generating functions \cite{Axt1998}:
\begin{subequations}\label{Eq_GeneratingFunctions}
\begin{align}
    Y(\alpha_{\bf q},\beta_{\bf q},t) &= \Braket{dc~\e^{\sum\limits_{\bf q}\alpha_{\bf q} b^\dagger_{\bf q}}\e^{\sum\limits_{\bf q} \beta_{\bf q} b_{\bf q}}},\\
    C(\alpha_{\bf q},\beta_{\bf q},t) &= \Braket{c^\dagger c~\e^{\sum\limits_{\bf q}\alpha_{\bf q} b^\dagger_{\bf q}}\e^{\sum\limits_{\bf q} \beta_{\bf q} b_{\bf q}}},\\
    F(\alpha_{\bf q},\beta_{\bf q},t) &= \Braket{\e^{\sum\limits_{\bf q}\alpha_{\bf q} b^\dagger_{\bf q}}\e^{\sum\limits_{\bf q} \beta_{\bf q} b_{\bf q}}},
\end{align}
\end{subequations}
where, for the purposes of deriving spectroscopic observables, the first function $Y$ returns the polarization:
\begin{align}
    P(t) &= \left.M^*Y(\alpha_{\bf q},\beta_{\bf q},t)\right|_{\alpha_{\bf q} = \beta_{\bf q} = 0},
\end{align}
while the other two functions $C$ and $F$ return the electronic population and bath amplitude respectively \cite{Vagov2002}.

In the following, we begin by reiterating the solution for excitation by a single pulse derived in \cite{Vagov2002}. We will then present the main result of this study, which is an expression for excitation by three pulses that describes multidimensional coherent spectroscopies. This expression will be demonstrated for two case studies, namely dynamics of polaron formation in electronic linewidths and coherent signatures of coupling to discrete bath modes.

\section{Single-Pulse Excitation}

\begin{figure}[H]
    \centering
    \includegraphics[width=0.45\textwidth]{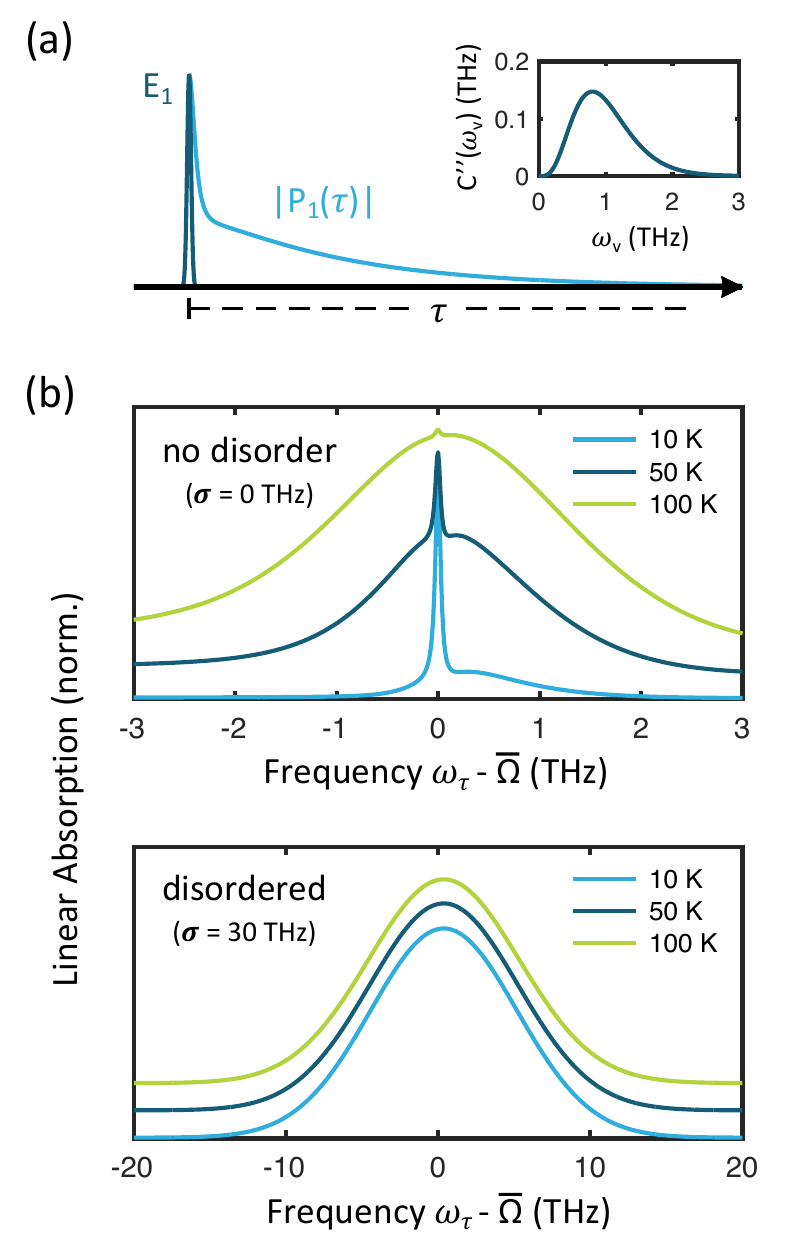}
    \caption{(a) Schematic of the single excitation pulse $E_1$ which impulsively excites a polarization $P_1(\tau)$, simulated for a temperature of 10 K and the parameters described in the text. (b) Corresponding absorption spectra for three temperatures as indicated. Simulations are performed for both no disorder (top panel) and a Gaussian resonance energy distribution of width $\sigma = 30$ THz.}
    \label{Fig1}
\end{figure}

We first consider the scenario of excitation by a single pulse of pulse area $f_1$ that arrives at $\tau = 0$ (see Fig.~\ref{Fig1}a). For initial conditions of the system in its ground state and the bath at thermal equilibrium, the resultant polarization $P_1(\tau)$ was derived in \cite{Vagov2002} as:
\begin{align}\label{Eq_P1}
    \nonumber P_1(\tau) &= \frac{i}{2}\sin(f_1)e^{-i\overline{\Omega}\tau}e^{i{\bf k}_1\cdot {\bf r}}e^{-\gamma \tau}\\
    &\hspace{0.4cm}\times \exp\Bigg[\int\frac{C''(\omega_v)}{\omega_v^2}\left\{(e^{-i\omega_v \tau} - 1) - N_v|(e^{-i\omega_v \tau} - 1)|^2\right\}d\omega_v\Bigg],
\end{align}
where $\overline{\Omega} = \Omega - \int \frac{C''(\omega_v)}{\omega_v^2}\omega_v d\omega_v$ incorporates the lattice reorganization energy, $\gamma$ is a phenomenological homogeneous dephasing rate, $N_v = 1/\left(\exp\left[\hbar\omega_v/k_B\mathcal{T}\right] - 1\right)$ is the thermal distribution of the bath at a temperature $\mathcal{T}$, and the dipole moment $M$ is set to unity for simplicity.

The Fourier transform of the temporal polarization $P_1(\omega_\tau)$ directly leads to a corresponding linear absorption spectrum \cite{HammZanni2012}:
\begin{align}
    A(\omega_\tau) \propto -\text{Re}\left\{iP_1(\omega_\tau)\right\},
\end{align}
simulated for three representative temperatures and plotted in the top panel of Fig.~\ref{Fig1}b. The absorption spectrum exhibits a complex lineshape at the lowest temperature of 10 K that becomes gradually featureless up to 70 K. Besides the sharp zero-phonon line \cite{Hsu1984} (broadened by $\gamma = 0.1$ THz), the surrounding pedestal physically corresponds to inelastic scattering with low-frequency vibrations and is therefore a direct characterization of the spectral density \cite{Krummheuer2002,Liu2019_JPCL}. In many realistic systems however, there is unavoidable disorder of the electronic resonance energy $\overline{\Omega}$ and the polarization must be convolved with the resonance energy distribution. Absorption spectra incorporating a Gaussian resonance energy distribution of width $\sigma = 30$ THz are plotted in the bottom panel of Fig.~\ref{Fig1}b, whose intricate homogeneous lineshapes have been completely obscured by the disorder. Resolving these lineshapes in the presence of such electronic disorder is a primary motivation for applying multidimensional spectroscopy, which will be the focus of the remaining text.

\section{Three-Pulse Excitation}

We now consider the general scenario of excitation by three pulses $\{E_1,E_2,E_3\}$, with inter-pulse time delays $\{\tau,T\}$ and an emission time $t$ (see Fig.~\ref{Fig2}a). By sequential propagation of each generating function (described in the Appendix), one obtains numerous polarization terms that emit in unique phase-matched directions. Here, we focus on deriving an expression for the photon-echo signal most commonly measured in MDCS experiments (emitted with a wavevector ${\bf k}_{sig} = -{\bf k}_1 + {\bf k}_2 + {\bf k_3}$), and arrive at the main result of this study:
\begin{widetext}
    \begin{align}\label{eq_P3}
        \nonumber P_3^{echo}(\tau,T,t) &= -\frac{i}{4}\sin(f_1)\sin(f_2)\sin(f_3)e^{i\overline{\Omega}(\tau - t)}\e^{-i({\bf k}_1 - {\bf k}_2 - {\bf k}_3)}e^{-\gamma(t + \tau)}\\
        \nonumber &\hspace{0.4cm}\times \exp\Bigg[\int\frac{C''(\omega_v)}{\omega_v^2}\left\{e^{-i\omega_v t} + e^{i\omega_v(T + \tau)} - e^{i\omega_v(\tau+T+t)} + e^{i\omega_v\tau} + e^{i\omega_v(T+t)} - e^{i\omega_vT} - 2\right.\\
        &\hspace{7cm}- \left.N_v|\e^{i\omega_v(\tau+T)} -e^{i\omega_v(\tau + T + t)} + \e^{i\omega_v\tau} - 1|^2\right\} d\omega_v\Bigg],
    \end{align}  
\end{widetext}

which is immediately seen to be more compact than the usual perturbative expression \cite{Butkus2012,Seibt2013,Seibt2013_2}. We point out two additional advantages of this expression. First, that this expression simplifies interpretation by separating the zero-temperature dynamics of the coupled system and bath (first bracketed term) from the stimulated dynamics induced by thermal occupation of the bath (second bracketed term). Second, that corresponding dynamics of the electronic and vibrational populations can be directly obtained from the generating functions $C$ and $F$ \cite{Vagov2002}. In addition to this rephasing term of the nonlinear polarization, a corresponding non-rephasing polarization (emitted with a wavevector ${\bf k}_{sig} = {\bf k}_1 - {\bf k}_2 + {\bf k}_3$) is often desired as well, for example to generate absorptive 2-D spectra \cite{HammZanni2012}. This non-rephasing polarization may be found in an analogous fashion, and is presented in the Appendix. We perform demonstrative simulations of expression (\ref{eq_P3}) in the following.

As shown in Fig.~\ref{Fig2}a, the polarization $|P_3^{echo}(\tau,T,t)|$ exhibits a characteristic revival at $t = \tau$ due to `rephasing' of the disordered Bloch vectors \cite{MDCS_Book}, whose width is the inverse of the disorder linewidth $\sigma$. By integrating (\ref{eq_P3}) along the emission time $t$, one can describe integrated photon-echo experiments as shown in Fig.~\ref{Fig2}b (simulated for the same Debye spectral density and disorder linewidth $\sigma$ as above). As observed in experiments, such as on quantum dots \cite{Masia2012,Becker2018}, quantum wells \cite{Langer2014}, and 2-D van der Waals materials \cite{Dey2016}, the dephasing along $\tau$ exhibits both slow and fast decay components, corresponding to dephasing of the zero-phonon line and from bath-induced scattering respectively. With increasing temperature the fast decay component dominates, corresponding to the growth of the pedestal in Fig.~\ref{Fig1}b.

The simulations in Fig.~\ref{Fig2}b demonstrate how integrated photon-echo spectroscopy is able to resolve homogeneous dephasing dynamics in the presence of disorder. However, this technique suffers from two primary limitations. First, with the goal of characterizing the spectral density in mind, subtle differences in the bath-induced fast decay component are difficult to interpret in the time-domain. Second, the dephasing dynamics probed by integrated techniques necessarily measure an ensemble-averaged quantity, which can obscure important variations across the disordered ensemble. Both of these limitations may be circumvented by spectrally-resolving the nonlinear absorption and emission processes as well as their correlations, a primary capability of MDCS.

To correlate absorption and emission by MDCS, one measures the polarization $P^{echo}_3$ along two time delays $\{\tau,t\}$ and simultaneous Fourier transform along both time axes returns a two-dimensional (2-D) spectrum. 2-D spectra simulated via (\ref{eq_P3}) are shown in Fig.~\ref{Fig2}c, in which the vertical `absorption' and horizontal `emission' frequency axes correspond to the Fourier conjugate axes of $\tau$ and $t$ respectively. The utility of such a 2-D spectrum comes from the fact that the disorder lineshape and the homogeneous spectral lineshape are projected into orthogonal directions \cite{Siemens2010}, in principle permitting a measurement of the spectral density across the resonance frequency distribution \cite{Liu2021_ACSNano}.

To illustrate this idea, we take a vertical slice of the spectrum at 10 K to obtain a pseudo-absorption spectrum, and plot this in Fig.~\ref{Fig2}d with the linear absorption spectrum in Fig.~\ref{Fig1}b (with no disorder) for comparison. Besides a clear correspondence between the two lineshapes, the pedestal feature is notably more prominent in the slice which reflects a {\it nonlinear enhancement} of system-bath coupling signatures in 2-D spectra. Indeed, MDCS was recently applied to colloidal CdSe quantum dots at cryogenic temperatures for this purpose \cite{Liu2019_JPCL}, demonstrating the technique as a direct, sensitive probe of the spectral density of system-bath coupling.

\section{Two Case Studies}

We now demonstrate the utility of expression (\ref{eq_P3}) by reproducing observables in two exemplary experiments, namely in studying polaron formation through 2-D lineshape analysis, and resolving coherent phonon dynamics using MDCS. Both of these case studies are performed on excitons in a host crystal lattice, but have direct analogues to any generic open quantum system characterized by a spectral density with a low-frequency continuum and discrete high-frequency mode respectively.

\subsection{Polaron Formation Dynamics}

The concept of a non-equilibrium bath amplitude following excitation of a system is ubiquitous. In molecular systems this is clearly represented in a displaced oscillator model \cite{Nitzan2006}, in which transition to an excited state potential changes the bath energy minimum. In condensed systems (with a lattice) a continuum of bath modes of varying momenta are simultaneously displaced, resulting in a standing wave surrounding the excitation. The resulting quasiparticle incorporating a lattice potential well is known as a polaron \cite{Emin2013_Book}, a concept which has proved central to understanding a wide swath of functional materials \cite{Franchini2021}.

In a recent study by Seiler et al. \cite{Seiler2019}, MDCS was applied to CsPbI$_3$ nanocrystals towards resolving polaron formation dynamics. Although the experiment was performed at room temperature, where exciton lineshapes are relatively featureless with the linewidth being its sole characteristic, polaron formation may be inferred by tracking the linewidth as a function of intermediate time-delay $T$. With increasing $T$, absorption and emission frequencies become increasingly uncorrelated due to lattice relaxation, resulting in an effective broadening of the exciton lineshape \cite{Singh2016}.

In Fig.~\ref{Fig3}a, we plot a 2-D spectrum at room temperature and extract the resonance linewidth from vertical slices taken along the blue arrow indicated. The linewidth is observed to vary as a function of $T$, which is quantified in Fig.~\ref{Fig3}b with the full-width at half max as a function of $T$. The linewidth increases rapidly in the first 400 fs, very similar to the behavior attributed to polaron formation observed in \cite{Seiler2019}. We note that at low temperatures, polaron formation additionally manifests in evolution of the pedestal feature around the zero-phonon line, becoming more symmetric as vibrational absorption and emission processes equilibrate in the excited state potential \cite{Krugel2007,Wigger2020}.

\subsection{Coherent Phonon Dynamics}

Up to this point, we have considered only a Debye contribution to the spectral density that results in overdamped, incoherent dynamics. However, discrete molecular or lattice vibrations may also be incorporated into the spectral density, for example through a Lorentzian contribution of the form \cite{Butkus2012,Seibt2013}:
\begin{align} \label{Eq_SpectralDensity_Lorentzian}
    C''_L(\omega_v) = \frac{2^{3/2}S_L\omega_L^3\gamma_L\omega_v}{(\omega^2_v - \omega_L^2)^2 + 2\gamma_L^2\omega_v^2},
\end{align}

where we use the values for the Huang-Rhys factor $S_L = 0.05$, the mode frequency $\omega_L = 6$ THz, and a damping rate $\gamma_L = 0.75$ THz. These values are chosen to mimic exciton coupling to the longitudinal-optical (LO) phonon in CdSe, a simple model system that has been extensively studied \cite{Kelley2019}. 

Exciton coupling to LO phonons in CdSe quantum dots was studied by us \cite{Liu2019_PRL} and others \cite{Caram2014,Wang2017,Gellen2017,Wang2022} using MDCS, in which LO phonons manifested as clear sidebands adjacent to the exciton peak in the 2-D spectra. A nonlinear analogue to phonon replicas \cite{Kelley2019,Liu2023_ES} in linear absorption or luminescence spectra, these sidebands additional exhibit coherent dynamics that provide direct insight into the microscopic parameters of exciton-phonon coupling. 

In Fig.~\ref{Fig4}a, we present 2-D spectra simulated at 10 K which incorporate both the Debye contribution to the spectral density in (\ref{Eq_SpectralDensity}) as well as the Lorentzian contribution in (\ref{Eq_SpectralDensity_Lorentzian}). In agreement with experiment, two sidebands are observed above and below the primary exciton peak whose ratio depends on the intermediate delay $T$. To examine the dynamics more closely, slices are taken along the blue arrow in (a) as a function of $T$ and plotted in Fig.~\ref{Fig4}b. Indeed, strong coherent oscillations are observed of the lower sideband in strong agreement with experiment. We emphasize again that the above simulations and referenced experiments are all performed in the presence of a large distribution of resonance energies, underscoring the powerful ability of MDCS to extract information about the spectral density even in the presence of dominant disorder broadening.

\section{Conclusion}

In summary, we have presented an exact and compact expression for multi-dimensional spectroscopy of a spin-boson Hamiltonian. The expression was derived from a generating function approach by Vagov et al. \cite{Vagov2002}, originally developed for linear and integrated two-pulse four-wave mixing experiments. We have demonstrated its validity in two case studies from literature that study continuum low-frequency and discrete high-frequency components of the spectral density respectively. These are a study of polaron formation dynamics through an ultrafast increase in transition linewidth \cite{Seiler2019}, and a study of coherent exciton-phonon coupling dynamics \cite{Caram2014,Wang2017,Gellen2017,Liu2019_PRL,Wang2022} by MDCS. The expression derived here successfully reproduces both experimental observables.

There are of course many systems that require a more sophisticated approach to model, for example those with multiple participating quantum states \cite{Turner2010,Luttig2023}, higher-order system-bath couplings \cite{Loring2022}, or a bath with strong anharmonicities \cite{Okumura1996}. Excitation schemes that use pulses which break the impulsive approximation also will require numerical approaches. Nevertheless, the expression presented here is expected to apply to a broad range of systems, and greatly simplify simulating MDCS of open quantum systems.

\section{Acknowledgments}

A.L. was supported by the U.S. Department of Energy, Office of Basic Energy Sciences, under Contract No. DE-SC0012704.

\newpage
\begin{figure}[H]
    \centering
    \includegraphics[width=0.45\textwidth]{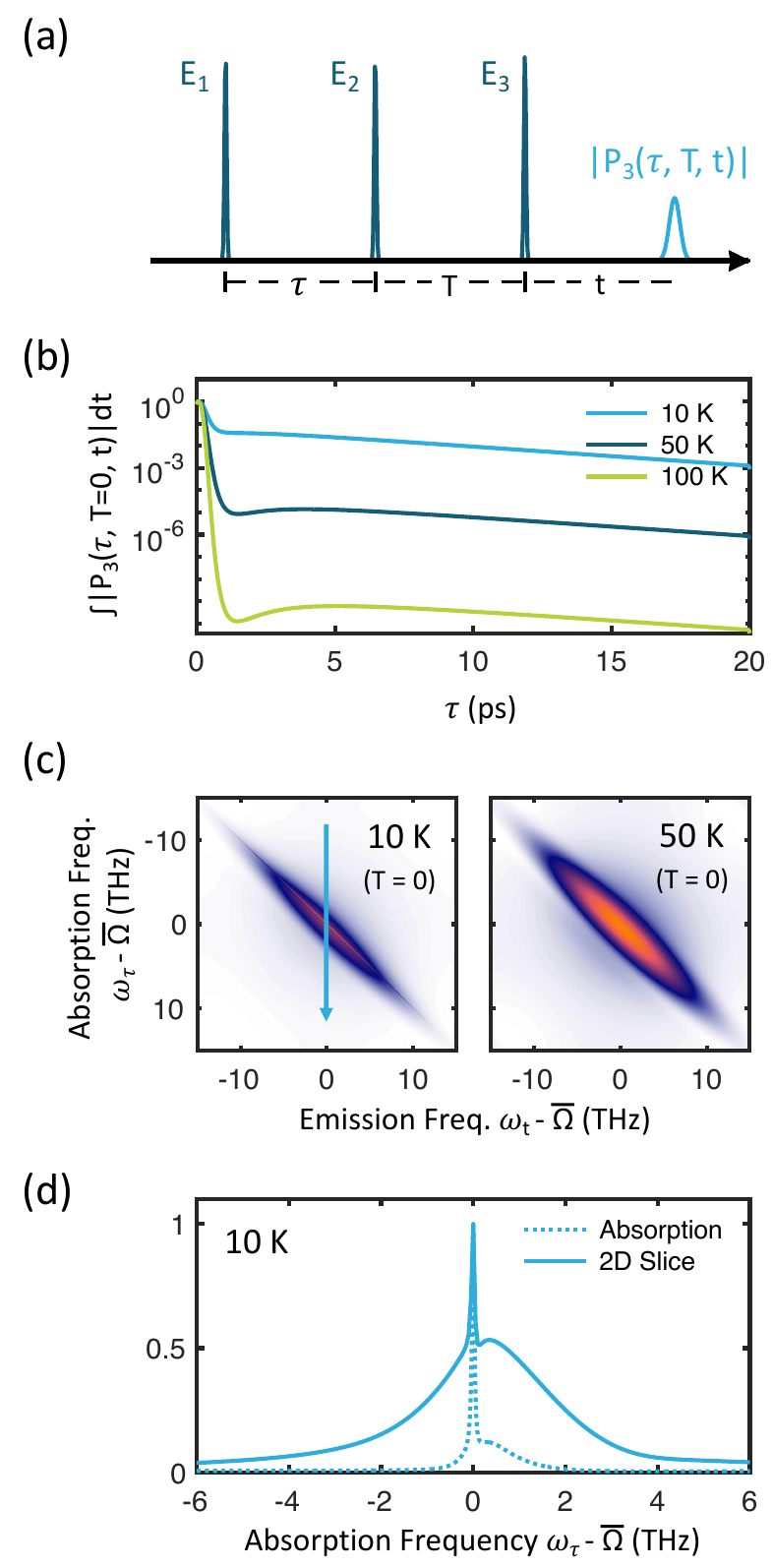}
    \caption{(a) Schematic of the three excitation pulses \{$E_1,E_2,E_3$\} which impulsively excite a polarization $P_3(\tau,T,t)$. (b, c) Simulations of (b) integrated photon-echo and (c) MDCS performed for a disorder linewidth of $\sigma = 30$ THz and $T = 0$. (d) Comparison at 10 K of an absorption spectrum (no disorder) to the 2-D spectrum slice taken along the blue arrow in (c). A nonlinear enhancement of the pedestal, arising from system-bath coupling, is observed in the nonlinear 2-D spectrum.}
    \label{Fig2}
\end{figure}
\newpage
\begin{figure}[H]
    \centering
    \includegraphics[width=0.45\textwidth]{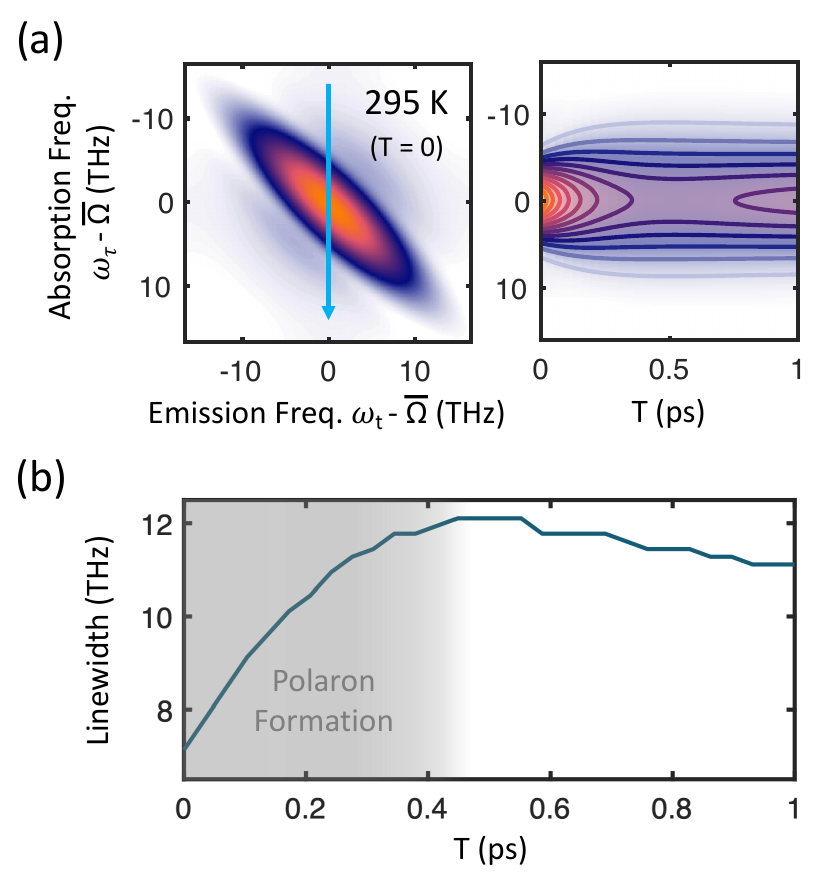}
    \caption{(a) Simulated 2-D spectrum at room temperature and $T = 0$. Blue arrow indicates position of slices shown in the right panel as a function of intermediate delay $T$. (b) Full-width at half max extracted from slices in (a) which increases rapidly in the first $\approx$400 fs, indicating polaron formation.}
    \label{Fig3}
\end{figure}
\newpage
\begin{figure}[H]
    \centering
    \includegraphics[width=0.5\textwidth]{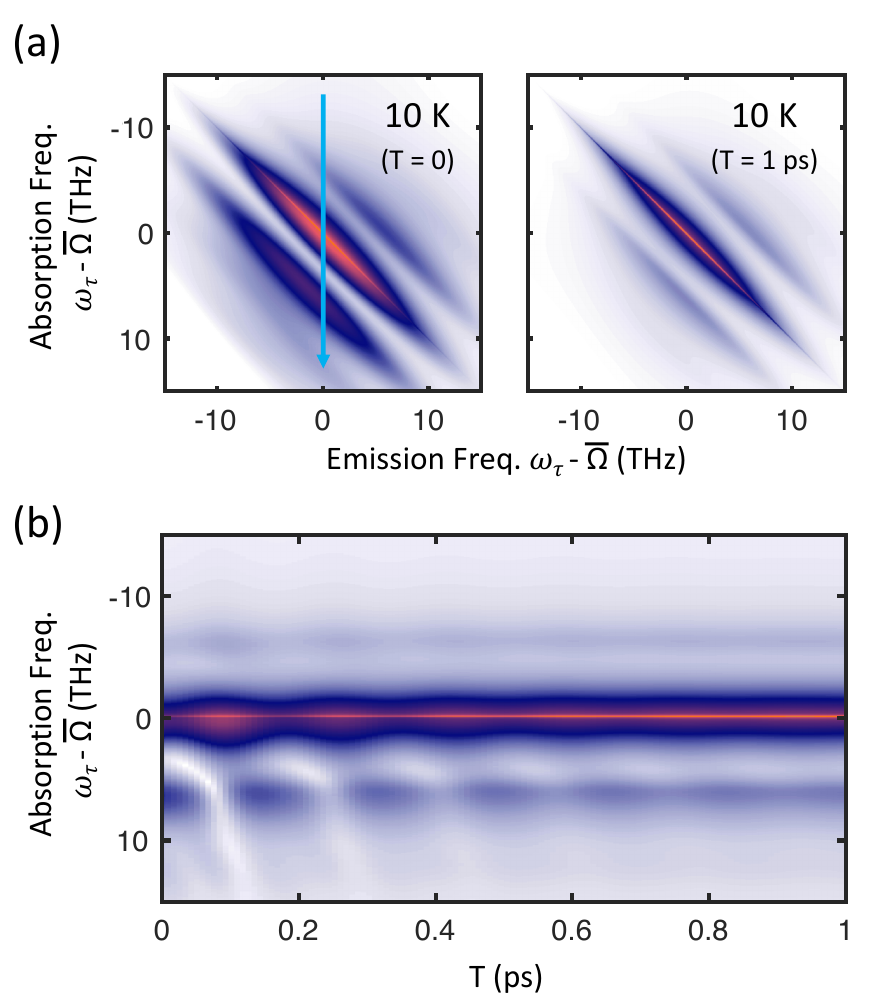}
    \caption{(a) 2-D spectra simulated for a spectral density incorporating a discrete phonon mode. The blue arrow indicates the location of the slices plotted in (b). (b) Dependence of the vertical slice indicated in (a) as a function of intermediate delay $T$. Strong coherent oscillations of the lower sideband are observed with minimal modulation of the upper sideband.}
    \label{Fig4}
\end{figure}

\newpage

\section{Appendix}

In this Appendix we describe the essential ingredients for deriving (\ref{eq_P3}), and for a more complete description of the theoretical framework we refer the reader to \cite{Vagov2002}. 

\subsection{Transformed Generating Functions}

As described in the text, the dynamics of both the system and bath are captured by the generating functions in (\ref{Eq_GeneratingFunctions}). For calculating the dynamics, it is convenient to transform these variables according to:
\begin{subequations}
\begin{align}
    \overline{Y}(\alpha_{\bf q},t) &= e^{i\overline{\Omega}t}\exp\left[\sum\limits_{\bf q}\gamma^*_{\bf q}\alpha_{\bf q} e^{i\omega_{\bf q} t}\right]Y\left(\alpha_{\bf q} e^{i\omega_{\bf q}t} - \gamma_{\bf q},t\right)\\
    \overline{C}(\alpha_{\bf q},t) &= \exp\left[2i\imag\left\{\sum\limits_{\bf q}\gamma^*_{\bf q}\alpha_{\bf q} e^{i\omega_{\bf q} t}\right\}\right]C\left(\alpha_{\bf q} e^{i\omega_{\bf q} t},t\right)\\
    \overline{F}(\alpha_{\bf q},t) &= F\left(\alpha_{\bf q}e^{i\omega_{\bf q}t},t\right)
\end{align}
\end{subequations}
where we define $\gamma_{\bf q} = (g^e_{\bf q} - g^h_{\bf q})/\omega_{\bf q}$ and use the shorthand $Y(\alpha_{\bf q},t) = Y(-\alpha_{\bf q}^*,\alpha_{\bf q},t)$, $C(\alpha_{\bf q},t) = C(-\alpha_{\bf q}^*,\alpha_{\bf q},t)$, and $F(\alpha_{\bf q},t) = F(-\alpha_{\bf q}^*,\alpha_{\bf q},t)$. The inverse transformations are then:
\begin{subequations}\label{eq_inversetransformation}
\begin{align}
    Y(\alpha_{\bf q},t) &= e^{-i\overline{\Omega}t}\exp\left[-\sum\limits_{\bf q}\gamma^*_{\bf q}(\alpha_{\bf q} + \gamma_{\bf q})\right]\overline{Y}\left(e^{-i\omega_{\bf q}t}(\alpha_{\bf q} + \gamma_{\bf q}),t\right)\\
    C(\alpha_{\bf q},t) &= \exp\left[-2i\imag\left\{\sum\limits_{\bf q}\gamma^*_{\bf q}\alpha_{\bf q}\right\}\right]\overline{C}\left(\alpha_{\bf q}e^{-i\omega_{\bf q} t},t\right)\\
    F(\alpha_{\bf q},t) &= \overline{F}\left(e^{-i\omega_{\bf q}t}\alpha_{\bf q},t\right)
\end{align}
\end{subequations}
To obtain an analytical solution to the equations of motion for $\overline{Y}$, $\overline{C}$, and $\overline{F}$ \cite{Vagov2002}, we take the special case of impulsive excitation by infinitely short pulses. 

\subsection{Time Evolution of the Generating Functions}

\begin{figure}[H]
    \centering
    \includegraphics[width=0.45\textwidth]{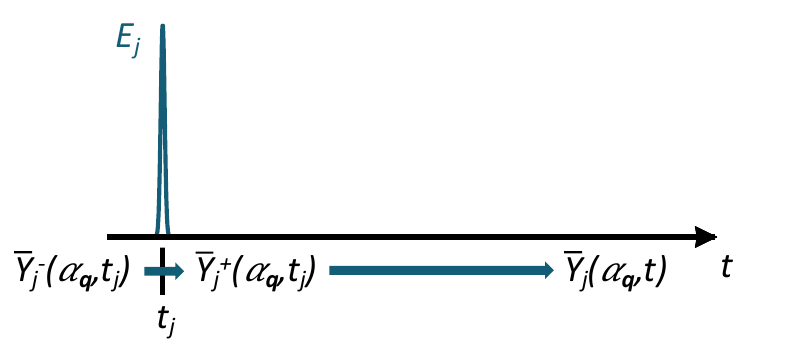}
    \caption{Schematic of the dynamics induced by an impulsive excitation $E_j$ on the generating function $\overline{Y}$. The values of $\overline{Y}$ immediately before and after excitation are denoted $\overline{Y}^-_j(\alpha_{\bf q},t_j)$ and $\overline{Y}^+_j(\alpha_{\bf q},t_j)$, and the latter evolves in time to a value $\overline{Y}_j(\alpha_{\bf q},t)$.}
    \label{FigA1}
\end{figure}

In this case, the time-evolution of the generating functions can be separated between three distinct quantities relative to an excitation pulse $E_j$. As illustrated in Fig.~\ref{FigA1} for the function $\overline{Y}$, we define $\overline{Y}^-_j(\alpha_{\bf q},t_j)$ and $\overline{Y}^+_j(\alpha_{\bf q},t_j)$ as the instantaneous values of $\overline{Y}$ directly before and after impulsive excitation by $E_j$ respectively. The quantity $\overline{Y}^+_j(\alpha_{\bf q},t_j)$ then evolves for a time period $t - t_j$ to a value $\overline{Y}_j(\alpha_{\bf q},t)$. The other two generation functions $\overline{C}$ and $\overline{F}$ may then be considered in the same way.

First, the time evolutions of $\overline{Y}$ and $\overline{C}$ are given simply by $\overline{Y}_j(\alpha_{\bf q},t) = \overline{Y}^+_j(\alpha_{\bf q},t_j)$ and $\overline{C}_j(\alpha_{\bf q},t) = \overline{C}^+_j(\alpha_{\bf q},t_j)$. The time evolution of $\overline{F}$ is then given by:
\begin{align}
    \nonumber \overline{F}_j(\alpha_{\bf q},t) &= \overline{F}_j^+(\alpha_{\bf q},t_j)\\
    \nonumber &\hspace{0.4cm} + \overline{C}^+_j(\alpha_{\bf q},t_j)\left(\exp\left[-2i\imag\left\{\sum\limits_{\bf q}\gamma^*_{\bf q}\alpha_{\bf q}e^{i\omega_{\bf q}t}\right\}\right] - \exp\left[-2i\imag\left\{\sum\limits_{\bf q}\gamma^*_{\bf q}\alpha_{\bf q}e^{i\omega_{\bf q}t_j}\right\}\right]\right)
\end{align}

This shows that while $\overline{F}$ evolves under the influence of an electronic population, $\overline{Y}$ and $\overline{C}$ remain static without external driving. The largest changes in the generating functions therefore occur immediately before and after $E_j$. These values are related by the following recursion relations:
\begin{widetext}
\begin{subequations}
\begin{align}
    \nonumber \overline{Y}_j^+(\alpha_{\bf q},t_j) &= \frac{1}{2}[1 + \cos(f_j)]\overline{Y}^-_j(\alpha_{\bf q},t_j) + \sin^2\left(\frac{f_j}{2}\right)\overline{Y}^{-*}_j(2\gamma_{\bf q}\e^{-i\omega_{\bf q} t_j} - \alpha_{\bf q},t_j)\e^{2i\phi_j + 2\sum\limits_{\bf q}\text{Re}\left[\gamma^*_{\bf q}(\alpha_{\bf q}\e^{i\omega_{\bf q} t_j} - \gamma_{\bf q})\right]}\\
    \nonumber &\hspace{0.4cm}+ \frac{i}{2}\sin(f_j)\e^{i\phi_j + \sum\limits_{\bf q}\gamma^*_{\bf q}\alpha_{\bf q}\e^{i\omega_{\bf q} t_j}}\Bigg[\overline{F}_j^-(\alpha_{\bf q} - \gamma_{\bf q}\e^{-i\omega_{\bf q} t_j},t_j) - \overline{C}_j^-(\alpha_{\bf q} - \gamma_{\bf q}\e^{-i\omega_{\bf q} t_j},t_j)\\
    &\hspace{4cm}\times \left.\left(e^{-2i\sum\limits_{\bf q}\text{Im}[\gamma^*_{\bf q}\alpha_{\bf q}\e^{i\omega_{\bf q} t_j}]} + e^{2i\sum\limits_{\bf q}\left\{|\gamma_{\bf q}|^2\sin[\omega_{\bf q}(t_{j-1} - t_j)] - \text{Im}[\gamma^*_{\bf q}\alpha_{\bf q}\e^{i\omega_{\bf q} t_{j-1}}]\right\}}\right)\right]\\
    \nonumber \overline{C}_j^+(\alpha_{\bf q},t_j) &= \overline{C}_j^-(\alpha_{\bf q},t_j) + \sin^2\left(\frac{f_j}{2}\right)\e^{2i\sum\limits_{\bf q}\text{Im}[\gamma^*_{\bf q}\alpha_{\bf q}\e^{i\omega_{\bf q} t_j}]}\Bigg[\overline{F}_j^-(\alpha_{\bf q},t_j)\\ 
    \nonumber &\hspace{5.5cm}\left. - \overline{C}_j^-(\alpha_{\bf q},t_j)\left(e^{-2i\sum\limits_{\bf q}\text{Im}[\gamma^*_{\bf q}\alpha_{\bf q} \e^{i\omega_{\bf q} t_j}]}+ e^{-2i\sum\limits_{\bf q}\text{Im}[\gamma^*_{\bf q}\alpha_{\bf q} \e^{i\omega_{\bf q} t_{j-1}}]}\right)\right]\\
    \nonumber &\hspace{0.4cm}- \frac{i}{2}\sin(f_j)\e^{2i\sum\limits_{\bf q}\text{Im}[\gamma^*_{\bf q}\alpha_{\bf q}\e^{i\omega_{\bf q} t_j}]}\left[\overline{Y}_j^-(\alpha_{\bf q} + \gamma_{\bf q}\e^{-i\omega_{\bf q} t_j},t_j)\e^{-i\phi_j - \sum\limits_{\bf q}[|\gamma_{\bf q}|^2 + \gamma_{\bf q}^*\alpha_{\bf q}\e^{i\omega_{\bf q} t_j}]}\right.\\ 
    &\hspace{6.8cm}\left.- \overline{Y}^{-*}_j(-\alpha_{\bf q} + \gamma_{\bf q}\e^{-i\omega_{\bf q} t_j},t_j)\e^{i\phi_j - \sum\limits_{\bf q}[|\gamma_{\bf q}|^2 - \gamma_{\bf q}\alpha^*_{\bf q}\e^{-i\omega_{\bf q} t_j}]}\right]\\
    \overline{F}_j^+(\alpha_{\bf q},t_j) &= \overline{F}_j^-(\alpha_{\bf q},t_j) + \overline{C}_j^-(\alpha_{\bf q},t_j)\left(\e^{-2i\text{Im}\left[\sum\limits_{\bf q} \gamma^*_{\bf q} \alpha_{\bf q}\e^{i\omega_{\bf q} t_j}\right]} - \e^{-2i\text{Im}\left[\sum\limits_{\bf q} \gamma^*_{\bf q} \alpha_{\bf q}\e^{i\omega_{\bf q} t_{j-1}}\right]}\right)
\end{align}
\end{subequations}
\end{widetext}
where $\phi_j = \overline{\Omega}t_j + {\bf k}_j$ and $f_j$ is the pulse area of the $j$th excitation pulse.

These recursion relations crucially allow one to propagate $\overline{Y}$, $\overline{C}$, and $\overline{F}$ in time across an arbitrary number of impulsive excitations. In the following, we propagate the generating functions in response to one, two, and finally three excitation pulses to arrive at the expression presented in (\ref{eq_P3}).

\subsection{One-Pulse Excitation}

We take the initial condition of a ground state electronic population and a phonon bath at thermal equilibrium:
\begin{align}
    Y(t = 0) = 0,\hspace{0.5cm} C(t = 0) = 0,\hspace{0.5cm} F(t = 0) = \e^{\sum\limits_{\bf q}\alpha_{\bf q}\beta_{\bf q} N_{\bf q}}
\end{align}
Or for the transformed variables:
\begin{subequations}
\begin{align}
    \overline{Y}^-_1(\alpha_{\bf q},t_1=0) &= 0\\ 
    \overline{C}^-_1(\alpha_{\bf q},t_1 = 0) &= 0\\
    \overline{F}^-_1(\alpha_{\bf q},t_1 = 0) &= \e^{-\sum\limits_{\bf q}|\alpha_{\bf q}|^2 N_{\bf q}}
\end{align}    
\end{subequations}
where we've defined $t_1 = 0$ as the center of time. Using the recursion relations, we find the values immediately after excitation by the first pulse:
\begin{subequations}
\begin{align}
    \overline{Y}_1^+(\alpha_{\bf q}) &= \frac{i}{2}\sin(f_1)e^{i\phi_1}e^{\sum\limits_{\bf q}\gamma^*_{\bf q}\alpha_{\bf q}}e^{-\sum\limits_{\bf q}|\alpha_{\bf q} - \gamma_{\bf q}|^2N_{\bf q}}\\
    \overline{C}_1^+(\alpha_{\bf q}) &= \sin^2\left(\frac{f_1}{2}\right)e^{2i\sum\limits_{\bf q}\imag\{\gamma^*_{\bf q}\alpha_{\bf q}\}}e^{-\sum\limits_{\bf q}|\alpha_{\bf q}|^2N_{\bf q}}\\
    \overline{F}_1^+(\alpha_{\bf q}) &= e^{-\sum\limits_{\bf q}|\alpha_{\bf q}|^2N_{\bf q}}
\end{align}
\end{subequations}
and propagating forward along a time axis we call $\tau$ (in anticipation of the second pulse arrival):
\begin{subequations} \label{FirstPulse}
\begin{align}
    \overline{Y}_1(\alpha_{\bf q},\tau) &= \overline{Y}_1^+(\alpha_{\bf q})\\ 
    \overline{C}_1(\alpha_{\bf q},\tau) &= \overline{C}_1^+(\alpha_{\bf q})\\
    \overline{F}_1(\alpha_{\bf q},\tau) &= e^{-\sum\limits_{\bf q}|\alpha_{\bf q}|^2N_{\bf q}}\left(1 + \sin^2\left(\frac{f_1}{2}\right)\left[e^{2i\sum\limits_{\bf q}\imag\{\gamma^*_{\bf q}\alpha_{\bf q}(1 - e^{i\omega_{\bf q}\tau})\}} - 1\right]\right)
\end{align}
\end{subequations}


\subsection{Two-Pulse Excitation}

We now add another excitation pulse that arrives at $t_2 = \tau$, and use (\ref{FirstPulse}) directly in the recursion relations.
\begin{widetext}
\begin{align}
    \nonumber \overline{Y}_2^+(\alpha_{\bf q}) &= \frac{i}{4}[1 + \cos(f_2)]\sin(f_1)e^{i\phi_1}e^{\sum\limits_{\bf q}\gamma^*_{\bf q}\alpha_{\bf q}}e^{-\sum\limits_{\bf q}|\alpha_{\bf q} - \gamma_{\bf q}|^2N_{\bf q}} + \frac{i}{2}\sin(f_2)\e^{i\phi_2 + \sum\limits_{\bf q}\gamma^*_{\bf q}\alpha_{\bf q}\e^{i\omega_{\bf q}\tau}}e^{-\sum\limits_{\bf q}|\alpha_{\bf q} - \gamma_{\bf q}\e^{-i\omega_{\bf q}\tau}|^2N_{\bf q}}\\
    \nonumber &- \frac{i}{2}\sin(f_1)\sin^2\left(\frac{f_2}{2}\right)e^{i(2\phi_2 - \phi_1)}e^{\sum\limits_{\bf q}\gamma_{\bf q}(2\gamma^*_{\bf q}\e^{i\omega_{\bf q}\tau} - \alpha^*_{\bf q})}e^{-\sum\limits_{\bf q}|(2\gamma_{\bf q}\e^{-i\omega_{\bf q}\tau} - \alpha_{\bf q}) - \gamma_{\bf q}|^2N_{\bf q}}\e^{2\sum\limits_{\bf q}\text{Re}\left[\gamma^*_{\bf q}(\alpha_{\bf q}\e^{i\omega_{\bf q}\tau} - \gamma_{\bf q})\right]}\\
    &- i\sin^2\left(\frac{f_1}{2}\right)\sin(f_2)\e^{i\phi_2 + \sum\limits_{\bf q}\gamma^*_{\bf q}\alpha_{\bf q}\e^{i\omega_{\bf q}\tau}}e^{-\sum\limits_{\bf q}|\alpha_{\bf q} - \gamma_{\bf q}\e^{-i\omega_{\bf q}\tau}|^2N_{\bf q}}\\
    \nonumber \overline{C}_2^+(\alpha_{\bf q}) &= \sin^2\left(\frac{f_1}{2}\right)e^{2i\sum\limits_{\bf q}\imag\{\gamma^*_{\bf q}\alpha_{\bf q}\}}e^{-\sum\limits_{\bf q}|\alpha_{\bf q}|^2N_{\bf q}} - 2\sin^2\left(\frac{f_1}{2}\right)\sin^2\left(\frac{f_2}{2}\right)\e^{2i\sum\limits_{\bf q}\text{Im}\{\gamma^*_{\bf q}\alpha_{\bf q}\e^{i\omega_{\bf q}\tau}\}}e^{-\sum\limits_{\bf q}|\alpha_{\bf q}|^2N_{\bf q}} \\ 
    \nonumber &\hspace{0.4cm}+ \sin^2\left(\frac{f_2}{2}\right)\e^{2i\sum\limits_{\bf q}\text{Im}[\gamma^*_{\bf q}\alpha_{\bf q}\e^{i\omega_{\bf q}\tau}]}e^{-\sum\limits_{\bf q}|\alpha_{\bf q}|^2N_{\bf q}} + \frac{1}{4}\sin(f_1)\sin(f_2)e^{-\sum\limits_{\bf q}|\gamma_{\bf q}|^2}\e^{2i\sum\limits_{\bf q}\text{Im}[\gamma^*_{\bf q}\alpha_{\bf q}\e^{i\omega_{\bf q}\tau}]}\\
    \nonumber &\hspace{0.4cm}\times \left[e^{i(\phi_1-\phi_2)}e^{\sum\limits_{\bf q}|\gamma_{\bf q}|^2\e^{-i\omega_{\bf q}\tau}}e^{-\sum\limits_{\bf q}|\alpha_{\bf q} + \gamma_{\bf q}\e^{-i\omega_{\bf q}\tau} - \gamma_{\bf q}|^2N_{\bf q}}\e^{\sum\limits_{\bf q}\gamma_{\bf q}^*\alpha_{\bf q}(1 - \e^{i\omega_{\bf q}\tau})}\right.\\ 
    &\hspace{5cm}\left.+ e^{-i(\phi_1-\phi_2)}e^{\sum\limits_{\bf q}|\gamma_{\bf q}|^2\e^{i\omega_{\bf q}\tau}}e^{-\sum\limits_{\bf q}|-\alpha_{\bf q} + \gamma_{\bf q}\e^{-i\omega_{\bf q}\tau} - \gamma_{\bf q}|^2N_{\bf q}}\e^{\sum\limits_{\bf q}\gamma_{\bf q}\alpha^*_{\bf q}(\e^{-i\omega_{\bf q}\tau} - 1)}\right]\\
    \overline{F}_2^+(\alpha_{\bf q}) &= e^{-\sum\limits_{\bf q}|\alpha_{\bf q}|^2N_{\bf q}}\left(1 + 2\sin^2\left(\frac{f_1}{2}\right)\left[e^{2i\sum\limits_{\bf q}\imag\{\gamma^*_{\bf q}\alpha_{\bf q}(1 - e^{i\omega_{\bf q}\tau})\}} - 1\right]\right)
\end{align}
\end{widetext}
where in $\overline{Y}$ we now have different terms that correspond not only to the individual responses of $E_1$ and $E_2$, but also nonlinear mixing terms. We then propagate these values along a time axis we now call $T$:
\begin{subequations}
\begin{widetext}
    \begin{align}
        \overline{Y}_2(\alpha_{\bf q},\tau,T) &= \overline{Y}_2^+(\alpha_{\bf q})\\ 
        \overline{C}_2(\alpha_{\bf q},\tau,T) &= \overline{C}_2^+(\alpha_{\bf q})\\
        \nonumber \overline{F}_2(\alpha_{\bf q},\tau,T) &= e^{-\sum\limits_{\bf q}|\alpha_{\bf q}|^2N_{\bf q}}\left(1 + 2\sin^2\left(\frac{f_1}{2}\right)\left[e^{2i\sum\limits_{\bf q}\imag\{\gamma^*_{\bf q}\alpha_{\bf q}(1 - e^{i\omega_{\bf q}\tau})\}} - 1\right]\right)\\ 
        &\hspace{0.4cm}+ \overline{C}^+_2(\alpha_{\bf q})\left(e^{-2i\sum\limits_{\bf q}\imag\left\{\gamma^*_{\bf q}\alpha_{\bf q}e^{i\omega_{\bf q}(\tau+T)}\right\}} - e^{-2i\sum\limits_{\bf q}\imag\left\{\gamma^*_{\bf q}\alpha_{\bf q}e^{i\omega_{\bf q}\tau}\right\}}\right)
    \end{align}
\end{widetext}
\end{subequations}

\subsection{Three-Pulse Excitation}

We now add a final third pulse that arrives at $t_3 = \tau + T$. The recursion relations now return a huge number of terms for all three generating functions, in particular for $\overline{Y}$ that returns the polarization. Fortunately in MDCS experiments only two terms are typically measured, the `rephasing' term that results in a photon echo and its corresponding non-rephasing term. These emit with wavevectors ${\bf k}_{sig}^{(R)} = -{\bf k}_1 + {\bf k}_2 + {\bf k}_3$ and ${\bf k}_{sig}^{(NR)} = {\bf k}_1 - {\bf k}_2 + {\bf k}_3$ respectively, and we therefore need only calculate the terms in $\overline{Y}$ with the corresponding phase factors $e^{-i(\phi_1 - \phi_2 - \phi_3)}$ and $e^{i(\phi_1 - \phi_2 + \phi_3)}$ respectively:
\begin{widetext}
\begin{align}
    \nonumber \overline{Y}_3^{+(R)}(\alpha_{\bf q}) &= -\frac{i}{4}\sin(f_1)\sin(f_2)\sin(f_3)\e^{-i(\phi_1 - \phi_2 - \phi_3) + \sum\limits_{\bf q}\gamma^*_{\bf q}\alpha_{\bf q}\e^{i\omega_{\bf q}(\tau+T)}}\e^{\sum\limits_{\bf q}\gamma_{\bf q}\alpha_{\bf q}^*(\e^{-i\omega_{\bf q}\tau} - 1)}\\
    &\hspace{0.4cm}\times e^{\sum\limits_{\bf q}|\gamma_{\bf q}|^2(\e^{i\omega_{\bf q}(\tau +T)} + \e^{i\omega_{\bf q}\tau} - \e^{i\omega_{\bf q}T} - 1)}e^{-\sum\limits_{\bf q}|-\alpha_{\bf q} + \gamma_{\bf q}\e^{-i\omega_{\bf q}(\tau+T)} + \gamma_{\bf q}\e^{-i\omega_{\bf q}\tau} - \gamma_{\bf q}|^2N_{\bf q}}\\
    \nonumber \overline{Y}_3^{+(NR)}(\alpha_{\bf q}) &= -\frac{i}{4}\sin(f_1)\sin(f_2)\sin(f_3)\e^{i(\phi_1 - \phi_2 + \phi_3) + \sum\limits_{\bf q}\gamma^*_{\bf q}\alpha_{\bf q}(1 - \e^{i\omega_{\bf q}\tau} + \e^{i\omega_{\bf q}(\tau+T)})}\\
    &\hspace{0.4cm}\times e^{\sum\limits_{\bf q}|\gamma_{\bf q}|^2(\e^{-i\omega_{\bf q}T} + \e^{-i\omega_{\bf q}\tau} - \e^{-i\omega_{\bf q}(\tau+T)} - 1)}e^{-\sum\limits_{\bf q}|\alpha_{\bf q} - \gamma_{\bf q}\e^{-i\omega_{\bf q}(\tau+T)} + \gamma_{\bf q}\e^{-i\omega_{\bf q}\tau} - \gamma_{\bf q}|^2N_{\bf q}}
\end{align}
\end{widetext}

Transforming back to the original generating functions:
\begin{widetext}
\begin{align}
    \nonumber Y_3^{(R)}(\alpha_{\bf q}) &= -\frac{i}{4}\sin(f_1)\sin(f_2)\sin(f_3)e^{i\overline{\Omega}(\tau - t) -i({\bf k}_1 - {\bf k}_2 - {\bf k}_3)}\e^{\sum\limits_{\bf q}\gamma^*_{\bf q}(\alpha_{\bf q} + \gamma_{\bf q})(e^{-i\omega_{\bf q}t} - 1)}\e^{\sum\limits_{\bf q}\gamma_{\bf q}(\alpha_{\bf q}^* + \gamma_{\bf q}^*)(\e^{i\omega_{\bf q}(T + t)} - e^{i\omega_{\bf q}(\tau+T+t)})}\\
    &\hspace{0.4cm}\times e^{\sum\limits_{\bf q}|\gamma_{\bf q}|^2(\e^{i\omega_{\bf q}(\tau +T)} + \e^{i\omega_{\bf q}\tau} - \e^{i\omega_{\bf q}T} - 1)}e^{-\sum\limits_{\bf q}|-e^{-i\omega_{\bf q}(\tau+T+t)}(\alpha_{\bf q} + \gamma_{\bf q}) + \gamma_{\bf q}\e^{-i\omega_{\bf q}(\tau+T)} + \gamma_{\bf q}\e^{-i\omega_{\bf q}\tau} - \gamma_{\bf q}|^2N_{\bf q}}\\
    \nonumber Y_3^{(NR)}(\alpha_{\bf q}) &= -\frac{i}{4}\sin(f_1)\sin(f_2)\sin(f_3)\e^{-i\overline{\Omega}(\tau + t)  + i({\bf k}_1 - {\bf k}_2 + {\bf k}_3)}e^{\sum\limits_{\bf q}\gamma^*_{\bf q}(\alpha_{\bf q} + \gamma_{\bf q})(e^{-i\omega_{\bf q}(\tau+T+t)} - \e^{-i\omega_{\bf q}(T+t)} + \e^{-i\omega_{\bf q}t} - 1)}\\
    &\hspace{0.4cm}\times e^{\sum\limits_{\bf q}|\gamma_{\bf q}|^2(\e^{-i\omega_{\bf q}T} + \e^{-i\omega_{\bf q}\tau} - \e^{-i\omega_{\bf q}(\tau+T)} - 1)}e^{-\sum\limits_{\bf q}|e^{-i\omega_{\bf q}(\tau+T+t)}(\alpha_{\bf q} + \gamma_{\bf q}) - \gamma_{\bf q}\e^{-i\omega_{\bf q}(\tau+T)} + \gamma_{\bf q}\e^{-i\omega_{\bf q}\tau} - \gamma_{\bf q}|^2N_{\bf q}}
\end{align}
\end{widetext}

which gives the polarizations:
\begin{widetext}
\begin{align}
    P_3^{(R)}(\tau,T,t) &\propto -\frac{i}{4}\sin(f_1)\sin(f_2)\sin(f_3)e^{i\overline{\Omega}(\tau - t) -i({\bf k}_1 - {\bf k}_2 - {\bf k}_3)}e^{-\sum\limits_{\bf q}|\gamma_{\bf q}|^2|-e^{-i\omega_{\bf q}(\tau+T+t)} + \e^{-i\omega_{\bf q}(\tau+T)} + \e^{-i\omega_{\bf q}\tau} - 1|^2N_{\bf q}}\\
    \nonumber &\hspace{0.4cm}\times e^{\sum\limits_{\bf q}|\gamma_{\bf q}|^2(e^{-i\omega_{\bf q}t} + \e^{i\omega_{\bf q}(T + t)} - e^{i\omega_{\bf q}(\tau+T+t)} + \e^{i\omega_{\bf q}(\tau +T)} + \e^{i\omega_{\bf q}\tau} - \e^{i\omega_{\bf q}T} - 2)}\\
    \nonumber P_3^{(NR)}(\tau,T,t) &\propto -\frac{i}{4}\sin(f_1)\sin(f_2)\sin(f_3)\e^{-i\overline{\Omega}(\tau + t) + i({\bf k}_1 - {\bf k}_2 + {\bf k}_3)}e^{-\sum\limits_{\bf q}|\gamma_{\bf q}|^2|e^{-i\omega_{\bf q}(\tau+T+t)} - \e^{-i\omega_{\bf q}(\tau+T)} + \e^{-i\omega_{\bf q}\tau} - 1|^2N_{\bf q}}\\
    &\hspace{0.4cm}\times e^{\sum\limits_{\bf q}|\gamma_{\bf q}|^2(e^{-i\omega_{\bf q}(\tau+T+t)} - \e^{-i\omega_{\bf q}(T+t)} + \e^{-i\omega_{\bf q}t} + \e^{-i\omega_{\bf q}T} + \e^{-i\omega_{\bf q}\tau} - \e^{-i\omega_{\bf q}(\tau+T)} - 2)}
\end{align}
\end{widetext}
Finally, the summations over {\bf q} may be converted into integrals over bath frequency by use of the definition in (\ref{Eq_SpectralDensity}), to obtain the rephasing polarization as a function of spectral density expressed in (\ref{eq_P3}).

\end{document}